\documentclass{elsart}
\usepackage{amssymb}
\usepackage[pctex32]{graphics}
\usepackage{graphicx}
\graphicspath{{converted_graphics/}}

\begin{document}

\begin{frontmatter}

\title{Causality and the Entropy-Complexity Plane: \\ Robustness and
        Missing Ordinal Patterns}

\author[brazil,calculo,CONICET]{Osvaldo A. Rosso\corauthref{cor}},
\ead{oarosso@fibertel.com.ar}
\corauth[cor]{Corresponding author}
\author[newcastle,brazil]{Laura C. Carpi},
\ead{lauracarpi@gmail.com}
\author[newcastle]{Patricia M. Saco},
\ead{Patricia.Saco@newcastle.edu.au}
\author[brazil2]{Mart\'in G\'omez Ravetti},
\ead{martin.ravetti@dep.ufmg.br}
\author[iflp,CONICET]{Angelo Plastino}
\ead{plastino@fisica.unlp.edu.ar} and
\author[mdp,CONICET]{Hilda A. Larrondo}
\ead{larrondo@fi.mdp.edu.ar}

\address[brazil]{Departamento de F\'{\i}sica, Instituto de Ci\^encias Exatas.\\
                 Universidade Federal de Minas Gerais,\\
                 Av. Ant\^onio Carlos, 6627 - Campus Pampulha. \\
                 31270-901 Belo Horizonte - MG, Brazil.}

\address[calculo]{Chaos \& Biology Group,
                  Instituto de C\'alculo, \\
                  Facultad de Ciencias Exactas y Naturales.\\
                  Universidad de Buenos Aires.\\
                  Pabell\'on II, Ciudad Universitaria.\\
                  1428 Ciudad Aut\'onoma de Buenos Aires, Argentina.}

\address[newcastle]{Civil, Surveying and Environmental Engineering.\\
                    The University of Newcastle. \\
                    University Drive, Callaghan NSW 2308, Australia.}

\address[brazil2]{Departamento de Engenharia de Produ\c{c}\~ao, \\
                  Universidade Federal de Minas Gerais, \\
                  Av. Ant\^onio Carlos, 6627, Belo Horizonte, \\
                  31270-901 Belo Horizonte - MG, Brazil.}

\address[iflp]{Instituto de F\'isica, IFLP-CCT\\
             Universidad Nacional de La Plata (UNLP).\\
             C.C. 727, 1900 La Plata, Argentina.}

\address[mdp]{Facultad de Ingenier\'{\i}a, Universidad Nacional de Mar del Plata.\\
              Av. J.B. Justo 4302, 7600 Mar del Plata, Argentina}

\address[CONICET]{Fellow of CONICET-Argentina}

\begin{abstract}
We deal here with the issue of determinism versus randomness in
time series. One wishes to identify their relative weights in a
given time series. Two different tools have been advanced in the
literature to such effect, namely, {\it i)\/} the ``causal"
entropy-complexity plane [Rosso {\it et al.\/} Phys. Rev. Lett. 99
(2007) 154102] and {\it ii)\/} the estimation of the decay rate of
missing ordinal patterns [Amig\'o {\it et al.\/} Europhys. Lett.
79 (2007) 50001, and Carpi {\it et al.\/} Physica A 389 (2010)
2020–-2029]. In this work we extend the use of these techniques
to address the analysis of deterministic finite time series
contaminated
with additive noises of different degree of correlation. The
chaotic series studied here was via the logistic map ($r=4$) to
which we added correlated noise (colored noise with $f^{-k}$ Power
Spectrum, $0 \leq k \leq 2$) of varying amplitudes. In such a
fashion important insights pertaining to the deterministic
component of the original time series can be gained. We find that
in the entropy-complexity plane  this goal can be achieved
without additional computations.

PACS: 05.45.Tp; % (Time series analysis);
      02.50.-r; % (Probability theory, stochastic processes, and statistics)
      05.40.-a; % (Fluctuation phenomena, random processes, noise, and Brownian motion);
      05.40.Ca; % (Noise);
\vskip 0.5 mm
%Version: Physica A - V1
\end{abstract}

\maketitle
\end{frontmatter}

\section{Introduction}
\label{sec:Intro} The concept of
%%%%%%%%%%%%%%%%%OSVALDO 16-03-2011: agregue low dimensional
low dimensional
%%%%%%%%%%%%%%%%%%%%%%%%%%%%%%%%%%%%
deterministic chaos, derived from
the modern theory of nonlinear dynamical systems, has changed our
way of understanding and analyzing observational data $\mathcal
S(t)$ (time series), leading to a paradigm-change from linear to
non-linear approaches. Linear methods interpret  observational
signals from an underlying dynamical system that is regarded as
being  governed by a linear regime under which small perturbations
lead to  small effects. Consequently, all irregular behavior must
be  attributed to random external inputs \cite{Kantz2002}.
However, chaos theory has shown that random inputs are not the
only possible source of irregularities in a system's outputs. As a
matter of fact, nonlinear deterministic autonomous equations of
motion representing chaotic systems can give origin to very
irregular signals. Of course, a system which has both nonlinear
characteristics and random inputs will most likely produce
irregular signals as well \cite{Kantz2002}.

Chaotic time series are representative of a set of signals
exhibiting complex non-periodic traces with continuous, broad band
Fourier-spectra and displaying exponential sensitivity to small
changes in the initial conditions. Clearly, signals emerging from
chaotic time series occupy a place intermediate  between
{\it (a)\/} predictable regular or quasi-periodic signals and
{\it (b)\/} totally irregular
stochastic signals (noise)  which are completely unpredictable.
Chaotic time series are irregular in time, barely predictable, and
exhibit interesting  structures in the phase space.

Chaotic systems display  ``sensitivity to initial conditions"
which   manifests  instability everywhere in the phase space and
leads to  non-periodic motion (chaotic time series). They display
long-term unpredictability despite the deterministic character of
the temporal trajectory. In a system undergoing chaotic motion,
two neighboring points in the phase space move away exponentially
rapidly. Let ${\mathbf x}_1(t)$ and ${\mathbf x}_2(t)$ be two such
points,  located within a ball of radius $R$ at time $t$. Further,
assume that these two points cannot be resolved within the ball
due to poor instrumental resolution. At some later time $t'$ the
distance between the points will typically grow to $|{\mathbf
x}_1(t')-{\mathbf x}_2(t')| \approx |{\mathbf x}_1(t)-{\mathbf
x}_2(t)|~exp( \lambda |t'-t|)$, with $\lambda > 0$
for a chaotic dynamics and $\lambda$ the biggest Lyapunov
exponent. When this distance at time $t'$ exceeds $R$, the points
become experimentally distinguishable. This  implies that
instability reveals some information about the phase space
population that was not available at earlier times
\cite{Abarbanel1996}.

%%%%%%%%%%%%%%%%%%%%%%%%%%%%%
%%%%%%%%%%OSVALDO-13-03-2011: Aqui agregue el resumen de los metodos que nos manda a mirar el Referee 2
%%%%%%%%%%%%%%%%%%%%%%%%%%%%% Incorporadas las sugerencias de Angelo !!!!!
%%%%%%%%%%%%%%%%%%%%%%%%%%%%%
The above considerations allow one to think of chaos as an {\sl information source\/.}
Moreover, the associated rate of generated information can be formulated
in a precise way in terms of Kolmogorov-Sinai's entropy
\cite{Kolmogorov1958,Sinai1959}. The Kolmogorov-Sinai entropy
measures the average loss of information rate. Its range of values
goes from zero for regular dynamics, it is  a positive number for
chaotic system and, infinite for stochastic process. In
consequence, if a dynamical system exhibits at least one positive
Lyapunov exponent and a finite positive Kolmogorov-Sinai entropy
one can assert that the system is deterministic-chaotic.

Complex time series are very frequent in nature and also in
man-made systems. The immediate question that emerges in relation
to the underlying dynamical system that gives origin to these time
series reads: Is the system chaotic (low-dimensional
deterministic) or stochastic? Answering this question is important
for a proper physical description of irregular dynamics. If one is
able to show that the system is dominated by low-dimensional
deterministic chaos, then only few (nonlinear and collective)
modes are required to describe the pertinent dynamics
\cite{Osborne1989}. If not, then the complex behavior could be
modelled by a system dominated by a very large number of excited
modes which are in general better described by stochastic or
statistical approaches.

Even if several methodologies for evaluation of Lyapunov exponents
and Kol\-mogorov-Sinai entropies for time-series' analysis have been
proposed (see i.e \cite{Kantz2002}) their applicability involves
taking into account constraints (stationarity, time series length,
parameters values election for the methodology, etc.) which in
general make the ensuing results {\sl non-conclusive}. Thus, new
tools for distinguishing chaos (determinism) from noise
(stochastic) are needed.

In the  early days of chaotic dynamics it was believed that
obtaining finite, non-integer values for the fractal dimension was
a strong indication of the presence of deterministic chaos, in
opposition to systems whose dynamics are governed by stochastic
process,  thought to display an infinite value for the fractal
dimension. However, Osborne and Provenzale \cite{Osborne1989} were
able to present, in a seminal paper, a counter-example of the view
that for a  stochastic process one always detects a
non-convergence of the correlation dimension (as the estimation of
fractal dimension) in computed or measured time series. These
researchers show that curves (time series) generated by inverting
power-low spectra and random phases (colored random noises) are
random fractal paths. Their Housdorf dimension is finite, and thus
their correlation dimension is finite as well \cite{Osborne1989}.
The main reason is that the Grassberger and Procaccia approach for
the evaluation of the correlation dimension \cite{Grassberger1983}
is {\it independent of the points-ordering  in the signal\/} and
thus not able to test the differentiability of the curve under
study. Thus, the Grassberger and Procaccia method cannot
distinguish between fractal attractors and fractal curves if the
two have the same dimension \cite{Osborne1989}.

Among the many different tools advanced so as to distinguishing
chaotic from stochastic time series we can mention:
\begin{enumerate}
\item[{\it (a)}] The proposal of Sugihara and May \cite{Sugihara1990} based on
nonlinear forecasting. The main idea is to compare predicted and
actual trajectories. In this way, one can make tentative
distinctions between dynamical chaos and measurement errors. For a
chaotic time series the accuracy of nonlinear forecast diminishes
for increasing prediction time-intervals (at a rate which yields
an estimate of the Lyapunov exponent), whereas for uncorrelated
noise, the forecasting accuracy does not exhibit such dependency
\cite{Sugihara1990}.
\item[{\it (b)}] Kaplan and Glass
\cite{Kaplan1992,Kaplan1993} introduced a test based on the
observation that the tangent to the trajectory generated by a
deterministic system is a function of the position in phase space,
and therefore, all the tangents to a trajectory in a given phase
space region will display similar orientations, something that is
not observed in stochastic dynamics.
\end{enumerate}
Note that the two previously
mentioned methods have in common the fact that one has to choose a
certain length-scale $\epsilon$ and a particular embedding
dimension $D$.
\begin{enumerate}
\item[{\it (c)}] More recently, Kantz and co-workers
\cite{Kantz2000,Cecini2000} proposed to classify the signal
behavior, without referring to any specific model, as stochastic
or deterministic on a certain scale of resolution $\epsilon$,
according to the dependence of the $(\epsilon,\tau)$ entropy
$h(\epsilon,\tau)$ and finite-size Lyapunov exponent
$\lambda(\epsilon)$ on $\epsilon$. Their methodology for
distinguishing between chaos and noise is a refinement and
generalization of the Grassberger and Procaccia one
\cite{Grassberger1983} for estimating the correlation dimension,
and regarding  finite values as signatures of deterministic behavior.
\item[{\it (d)}] The use of quantifiers based in Information Theory which incorporate in their evaluation the ``time causality".
We must note, that in this methodology a length scale $\epsilon$ is chosen (that means the scale at which one measure the time series),
and also a particular embedding dimension $D$, however, its interpretation and main roll is completely different than in the previous
enumerates  methodologies (see for details see below).
\end{enumerate}
%%%%%%%%%%%%%%%%%%%%%%%%%%%%%%%%%%%%%%%%%%%%%%%%%%%%%%%%%%%%%%%%%%%%

%%%%%%%%%%%%OSVALDO 16-03-2011: frace de enganche con lo que teniamos antes
%onsequently, it is likely that the computation of quantifiers
%based on Information Theory concepts, like ``entropy",
%``statistical complexity", ``entropy-complexity plane", etc.,
The above mentioned quantifiers based on Information Theory
concepts, are (1) ``entropy", (2) ``statistical complexity" and
(3) the ``entropy-complexity plane". Judicious use of this plane
%%%%%%%%%%%%%%%%%%%%%%%%%%%%%%%
could {\it i)\/} lead to interesting insights into  the
characteristics of nonlinear chaotic dynamics, and {\it ii)\/}
potentially  improve  our understanding of their associated time
series. Moreover, these quantifiers can be used to detect
determinism in time series \cite{Rosso2007}. In this vein we
mention that Rosso {\it et al.\/} \cite{Rosso2007} found that
different Information Theory based measures (normalized Shannon
entropy and statistical complexity) allow for a better distinction
between deterministic chaotic and stochastic dynamics when the
so-called ``causal" information is incorporated via Bandt and
Pompe's (BP) methodology \cite{Bandt2002}. Indeed, new insight
into the characterization of theoretical and observational time
series, based on the Band and Pompe's methodology, reveals the
emergence of ``forbidden/missing patterns"
\cite{Amigo2006,Amigo2007,Amigo2008,Amigo2010}.  The BP approach
has been used also to distinguish deterministic behavior (chaos)
from randomness in finite time series contaminated with
%%%%%%%%%%%%OSVALDO 16-03-2011: aclaremos que es ruido blanco, ruido no correlacionado!!!!!
%observational noise
observational {\it white} noise (uncorrelated noise)
\cite{Amigo2006,Amigo2007,Carpi2010} by recourse to the analysis
of the decay rate of the ``missing ordinal patterns" as a function
of the time series length.

Additionally, Zanin \cite{Zanin2008} and Zunino {\it et al.\/}
\cite{Zunino2009} have recently studied the appearance of missing
ordinal patterns in financial time series. They found evidence for
the existence of deterministic forces in the medium- and long term
dynamics. Moreover, they propose that an analysis  of the number
of missing patterns' evolution should be a useful tool in order to
quantify the randomness of certain time-periods within a financial series.
Also, the presence of missing ordinal patterns has been
recently construed as
%%%%%%%%%%%%OSVALDO 16-03-2011: agregue "possible"
possible
%%%%%%%%%%%%%%%%%%%%%%%%%%%%%%%
evidence of deterministic dynamics in
epileptic states. It is suggested in \cite{Ouyang2009} that a
missing patterns' quantifier could be regarded as a predictor of
epileptic absence-seizures. It is of the essence to point out that
in such researches  only non correlated noise (white noise) has been
considered, which makes the associated  results somewhat
incomplete, since the  presence of colored noise has not been
taken into account.

It becomes necessary then, and such is the objective of this work,
to investigate from such viewpoint the robustness of the
entropy-complexity causality plane, that plays a prominent role in
some of the above cited discoveries. We intend to do this by
analyzing the planes's ability to distinguish between noiseless
chaotic time series and the ones that are contaminated with
additive correlated noise
%%%%%%%%%%%%OSVALDO 16-03-2011: aclaracion sobre el tipo de ruido
(noise with power low spectrum $f^{-k}$).
%%%%%%%%%%%%%%%%%%%%%%%%%%%%%%%

The chaotic series studied here were
generated by recourse to a logistic map to which noise with
varying amplitudes was added. The decay rate of missing ordinal
patterns was also investigated as a tool to distinguish among
these systems.

The present paper is organized as follows: Section
\ref{sec:Quantifiers} gives a brief description of the Information
Theory quantifiers used in this work. Section
\ref{sec:PDF-Bandt-Pompe}  presents the Bandt and Pompe
methodology used for the evaluation of the quantifiers. The
methodological framework used in this study is delineated in
Section \ref{sec:Logistica}. Finally Sections \ref{sec:Resultados}
and \ref{sec:Conclusions} present the discussion of the results
and the conclusions, respectively.

\section{Information Theory based quantifiers}
\label{sec:Quantifiers}

\subsection{Entropy and Statistical Complexity}
\label{sec:Quantifiers1}

The information content of a system is typically evaluated via a
probability distribution function (PDF) describing the
apportionment of some measurable or observable quantity. An
information measure can primarily be viewed as a quantity that
characterizes this given probability distribution $P$. The Shannon
entropy is very often used as a the ``natural" one
\cite{Shannon1949}. Given any arbitrary discrete probability
distribution $P = \{ p_i : i = 1, \cdots ,M \}$, with $M$ the
number of freedom-degrees,  Shannon's logarithmic information
measure reads
\begin{equation}
{\mathrm S}[P] = -\sum_{i=1}^{M}  p_i \ln( p_i) \ .
\label{Shannon}
\end{equation}
It can be regarded as a measure of the uncertainty associated to
the physical process described by $P$.
%%%%%%%%%OSVALDO: agregado aclaratorio para que nos rompan las bolls...
From now on we assume that the only restriction on the  PDF
representing the state of our system is $\sum_{j= 1}^N p_j = 1$
(micro-canonical representation).
%%%%%%%%%%%%%%%%
If ${\mathrm S}[P] = {\mathrm S}_{min} = 0$ we
are in position to predict with complete certainty which of the
possible outcomes $i$, whose probabilities are given by $p_i$,
will actually take place. Our knowledge of the underlying process
described by the probability distribution is maximal in this
instance. In contrast, our knowledge is minimal for a uniform
distribution and the uncertainty is maximal, ${\mathrm S}[P_e] =
{\mathrm S}_{max}$.

It is widely known that an entropic measure does not quantify the
degree of structure or patterns present in a process
\cite{Feldman1998}. Moreover, it was recently shown that measures
of statistical or structural complexity are necessary for a better
understanding of chaotic time series  because they are able
capture their organizational properties \cite{Feldman2008}. This
specific kind of information is not revealed by randomness'
measures. The opposite extreme perfect order (like a periodic
sequence) and maximal randomness (fair coin toss) possess no
complex structure and exhibit zero statistical complexity.
States between these extremes, a wide range of possible
degrees of physical structure exists, that should be quantified by
the {\it statistical complexity measure.\/} Rosso and coworkers
introduced an effective statistical complexity measure (SCM) that
is able to detect essential details of the dynamics and
differentiate different degrees of periodicity and chaos
\cite{Lamberti2004}. This specific SCM, abbreviated as the
 MPR one, provides important additional information regarding the
peculiarities of the underlying probability distribution, not
already detected by the entropy.

The MPR-statistical complexity measure is defined,  following the
seminal, intuitive notion advanced by L\'opez-Ruiz {\it et al.\/}
\cite{LopezRuiz1995}, via the product
\begin{equation}
{\mathcal C}_{JS}[P] = {\mathcal Q}_J [P, P_e] \cdot {\mathcal H}_S[P]
\label{Complexity}
\end{equation}
of i) the normalized Shannon entropy
\begin{equation}
{\mathcal H}_S[P] = {\mathrm S}[P] / {\mathrm S}_{max}  \ ,
\label{Shannon-normalizada}
\end{equation}
with ${\mathrm S}_{max} = {\mathrm S}[P_e]  = \ln M$, ($0 \leq
{\mathcal H}_S \leq 1$) and $P_e = \{ 1/M, \cdots , 1/M \}$ (the
uniform distribution) and ii) the so-called disequilibrium
${\mathcal Q}_J$. This quantifier is defined in terms of the
extensive (in the thermodynamical sense) Jensen-Shannon divergence
${\mathcal J} [P, P_e]$ that links two PDFs. We have
\begin{equation}
{\mathcal Q}_J [P, P_e] = Q_0 \cdot {\mathcal J} [P, P_e] \ ,
\label{Disequilibrium}
\end{equation}
with
\begin{equation}
{\mathcal J} [P, P_e] = {\mathrm S} \left[(P + P_e) / 2 \right] -
{\mathrm S}[P] / 2 - {\mathrm S}[P_e] / 2 \ .
\label{Jensen}
\end{equation}
 $Q_0$ is  a normalization constant, equal to the inverse of the
maximum possible value of ${\mathcal J} [P, P_e]$. This value is
obtained when one of the values of $P$, say $p_m$, is equal to one
and the remaining $p_i$ values are equal to zero, i.e.,
\begin{equation}
Q_0~=~-2 \left\{ \left( \frac{M+1}{M} \right) \ln(M+1) -2 \ln(2M) + \ln M \right\}^{-1} \ .
\label{Q0}
\end{equation}

The Jensen-Shannon divergence, that quantifies the difference
between two (or more) probability distributions, is especially
useful to compare the symbol-composition of different sequences
\cite{Grosse2002}. The complexity measure constructed in this way
has the intensive property found in many thermodynamic quantities
\cite{Lamberti2004}. We stress the fact that the statistical
complexity defined above is the product of two normalized
entropies (the Shannon entropy and Jensen-Shannon divergence), but
it is a nontrivial function of the entropy because it depends on
two different probabilities distributions, i.e., the one
corresponding to the state of the system, $P$, and the uniform
distribution, $P_e$.

\subsection{Entropy-Complexity plane}
\label{sec:Quantifiers2}

In statistical mechanics one is often interested in isolated
systems  characterized by an initial, arbitrary, and discrete
probability distribution. Evolution towards equilibrium is to be
described, as the overriding goal.
%%%%%%%OSVALDO: modifique
%The distribution corresponding to equilibrium conditions is the equiprobability distribution $P_e$.
At equilibrium, we can think, without loose of generality, that this state is given by the uniform distribution $P_e$.
%%%%%%%%%%%%%%%
The temporal evolution of the statistical complexity
measure (SCM) can be analyzed using a diagram of ${\mathcal
C}_{JS}$ versus time $t$. However, it is well known that the
second law of thermodynamics states that for isolated systems
entropy grows monotonically with time ($d{\mathcal H}_S/dt \geq
0$) \cite{Plastino1996}. This implies that ${\mathcal H}_S$ can be
regarded as an arrow of time, so that an equivalent way to study
the temporal evolution of the SCM is through the analysis of
${\mathcal C}_{JS}$ versus  ${\mathcal H}_S$. In this way, the
normalized entropy-axis substitutes for the time-axis.
Furthermore, it has been shown that for a given value of
${\mathcal H}_S$, the range of possible statistical complexity
values varies between a minimum ${\mathcal C}_{min}$ and a maximum
${\mathcal C}_{max}$ \cite{Martin2006},
restricting the possible values of SCM in this plane.

Therefore,  the evaluation of the complexity provides additional
insight into the details of the system's probability distribution,
which is not discriminated by randomness measures like the entropy
\cite{Rosso2007,Feldman2008}. It can also help to uncover
information related to the correlational structure between the
components of the physical process under study
\cite{Rosso2009B}.
The entropy-complexity diagram (or
plane), ${\mathcal H}_S \times {\mathcal C}_{JS}$, has been used
to study changes in the dynamics of a system originated by
modifications of some characteristic parameters
(see for instance Refs. \cite{Martin2006,Rosso2010,Zunino2010,Zunino2011} and references therein).

\subsection{Estimation of the Probability Distribution Function}
\label{sec:Quantifiers3}

In using  quantifiers based on Information Theory (like ${\mathcal
H}_S$ and ${\mathcal C}_{JS}$), a probability distribution
associated to the time series under analysis should be provided
beforehand. The determination of the most adequate PDF is a
fundamental problem because $P$ and the sample space $\Omega$ are
inextricably linked. Many methods have been proposed for a proper
selection of the probability space $(\Omega, P)$. We can mention:
{\it (a)\/} frequency counting \cite{Rosso2009C}, {\it (b)\/}
procedures based on amplitude statistics \cite{DeMicco2008}, {\it
(c)\/} binary symbolic dynamics \cite{Mischaikow1999}, {\it (d)\/}
Fourier analysis \cite{Powell979} and, {\it (e)\/} wavelet
transform \cite{Rosso2001}, among others. Their applicability
depends on particular characteristics of the data, such as
stationarity, time series length, variation of the parameters,
level of noise contamination, etc. In all these cases the
dynamics' global aspects can be somehow captured, but the
different approaches are not equivalent in their ability to
discern all the relevant physical details. One must also
acknowledge the fact that the above techniques are introduced in a
rather ``ad hoc fashion" and they are not directly derived from
the dynamical properties themselves of the system under study.

Methods for symbolic analysis of time  series that discretize the
raw  series and transform it into a  sequence of symbols
constitute a powerful tool. They efficiently analyze nonlinear
data and exhibit low sensitivity to noise \cite{Finn2003}.
However, finding a meaningful symbolic representation of the
original series is not an easy task \cite{Bollt2000,Daw2003}. To
our best knowledge, the Bandt and Pompe approach is the only
symbolization technique, among those in popular use, that takes
into account time-causality in the evaluation of the PDF
associated to the time series of a given system \cite{Bandt2002}.
The symbolic data is i) created by associating a rank to  the
series-values and ii) defined by reordering the embedded data in
ascending order. Data are reconstructed with an embedding
dimension $D$ (see definition and methodological details below).
In this way it is possible to quantify the diversity of the
ordering symbols (patterns) derived from a scalar time series,
evaluating the so called permutation entropy  ${\mathcal H}_s$,
and permutation statistical complexity ${\mathcal C}_{JS}$ (i.e.,
the normalized Shannon entropy and MPR-statistical complexity
evaluated for the Bandt and Pompe's PDF).

\section{The Bandt and Pompe methodology}
\label{sec:PDF-Bandt-Pompe}

\subsection{PDF based on ordinal patterns}
\label{sec:PDF-BP}

To use the Bandt and Pompe \cite{Bandt2002}  methodology for
evaluating of the probability distribution $P$ associated to the
time series (dynamical system) under study, one starts by
considering partitions of the pertinent $D$-dimensional space that
will hopefully ``reveal" relevant details of the ordinal-structure
of a given one-dimensional time series ${\mathcal S} = \{x_t :
t=1, \cdots, N \}$ with embedding dimension $D > 1$. We are
interested in ``ordinal patterns" of order $D$
\cite{Bandt2002,Keller2005} generated by
\begin{equation}
(s)~\mapsto~ \left(~x_{s-(D-1)},~x_{s-(D-2)},~\cdots,~x_{s-1},~x_{s}~\right) \ ,
\label{vectores}
\end{equation}
which assigns to each time $s$ the $D$-dimensional vector of values at times $s, s-1,\cdots,s-(D-1)$.
Clearly, the greater the $D-$value, the more information on the past  is incorporated into our vectors.
By ``ordinal pattern" related to the time $(s)$ we mean the permutation $\pi=(r_0,r_1, \cdots,r_{D-1})$ of $[0,1,\cdots,D-1]$ defined by
\begin{equation}
x_{s-r_{D-1}}~\le~x_{s-r_{D-2}}~\le~\cdots~\le~x_{s-r_{1}}~\le~x_{s-r_0} \ .
\label{permuta}
\end{equation}
In order to get a unique result we set $r_i <r_{i-1}$ if $x_{s-r_{i}}=x_{s-r_{i-1}}$.
This is justified if the values of $x_t$ have a continuous distribution so that equal values are very unusual.
Otherwise, it is possible to break these equalities by adding small random perturbations.

Thus, for all the $D!$ possible permutations $\pi$ of order $D$, their associated relative frequencies can be naturally computed by
the number of times this particular order sequence is found in the time series divided by the total number of sequences.
The probability distribution $P=\{p(\pi)\}$ is defined by
\begin{equation}
p(\pi)~=~\frac{\sharp \{s|s\leq M-D+1;~ (s) \texttt{, has type }\pi\}}{M-D+1} \ .
\label{frequ}
\end{equation}
In this expression, the symbol $\sharp$ stands for ``number".

The procedure can be better illustrated with a simple example; let
us assume that we start with the time series $\{1, 3, 5, 4, 2, 5,
\dots \}$, and we set the embedding dimension $D = 4$. In this
case the state space is divided into $4!$ partitions and $24$
mutually exclusive permutation symbols are considered. The first
4-dimensional vector is $(1, 3, 5, 4)$. According to
Eq.~(\ref{vectores}) this vector corresponds with $(x_{s-3},
x_{s-2}, x_{s-1}, x_{s})$. Following Eq.~(\ref{permuta}) we find
that $x_{s-3} \leq x_{s-2} \leq x{_s} \leq x_{s-1} $. Then, the
ordinal pattern which allows us to fulfill Eq.~(\ref{permuta})
will be
%%%%%%%%%%%%%OSVALDO 26-03-2011: el patron no es el que corresponde con la notacion
%$[0, 1, 3, 2]$.
$[3,2,0,1]$.
%%%%%%%%%%%%%%%%%%%%
The second 4-dimensional vector is $(3, 5, 4, 2)$, and
%%%%%%%%%%%%%OSVALDO: el patron no es el que corresponde con la notacion
%$[3, 0, 2, 1]$
$[0,3,1,2]$
%%%%%%%%%%%%%%%%%%%%
will be its associated permutation, and so on.
For the computation of the Bandt and Pompe PDF we follow the very fast algorithm described by Keller and Sinn \cite{Keller2005},
in which the different ordinal patterns are generated in lexicographic ordering.

The Bandt and Pompe's methodology is not restricted to  time series
representative of low dimensional dynamical systems but can be
applied to any type of time series (regular, chaotic, noisy, or
reality based), with a very weak stationary assumption (for $k =
D$, the probability for $x_t < x_{t+k}$ should not depend on $t$
\cite{Bandt2002}). It also assumes that enough data are available
for a correct embedding procedure. Of course, the embedding
dimension $D$ plays an important role in the evaluation of the
appropriate probability distribution because $D$ determines the
number of accessible states $D!$. It also conditions the minimum
acceptable length $M \gg D!$ of the time series that one needs in
order to work with reliable statistics.

The main difference between Information Theory quantifiers
evaluated with the Bandt and Pompe-PDF  and other PDFs (like a
histogram-PDF), is that they are invariant with respect to
strictly monotonous distortions of the data. This point is very
important for the analysis of observed natural time series (like
for example, sedimentary data related to rainfall by an unknown
nonlinear function \cite{Saco2010}). However, due to its
invariance properties, quantifiers based on the Bandt and Pompe's
approach, when  applied to any other series that is strictly
monotonic in the given data, will yield the same results.
Therefore, in general, ordinal data processing tools based on
ranked numbers  lead to results that
%%%%%%%%%OSVALDO: cambie la redaccion
%are more robust (and useful)
could be more useful
%%%%%%%%%%%%%%%%
for data analysis than those obtained using tools based on metric properties.

%%%%%%%%%%%OSVALDO-16-03-2011: movi esta subseccion aqui, a pedido del referee 3
\subsection{Forbidden and Missing Ordinal Patterns}
\label{sec:Quantifiers4}

As shown recently by Amig\'o {\it et al.\/}
\cite{Amigo2006,Amigo2007,Amigo2008,Amigo2010},  in the case of
deterministic one-dimensional maps, not all the possible ordinal patterns
%%%%%%%%%%%OSVALDO-16-03-2011: la aclaracion siguiente no es necesaria ahora !!!!
%(as defined below using Bandt and Pompe's methodology)
%%%%%%%%%%%%%%%%%%%%%%%%%%%%%
can be effectively materialized into orbits, which in a sense
makes these patterns ``forbidden". Indeed, the existence of these
{\it forbidden ordinal patterns\/} becomes a persistent fact that
can be regarded as a ``new" dynamical property. Thus, for a fixed
pattern-length (embedding dimension $D$) the number of forbidden
patterns of a time series (unobserved patterns) is independent of
the series length $N$. Remark that this independence does not
characterize other properties of the series such as proximity and
correlation, which die out with time \cite{Amigo2007,Amigo2010}.
For example, in the time series generated by the logistic map
$x_{k+1} = 4 x_k (1 - x_k)$, if we consider patterns of length
$D=3$, the pattern $\{2,1,0\}$ is forbidden. That is, the pattern
$x_{k+2} < x_{k+1} < x_{k}$ never appears \cite{Amigo2007}.

%%%%%%%%%%%%%%%%OSVALDO-17-03-2011: Mandamos al frente a Amigo !!!!
%%%%%%%%%%%%%%%%Osvaldo: correcciones de Angelo incorporadas
Stochastic process could also present forbidden patterns \cite{Golden}.
However, in the case of either uncorrelated or correlated stochastic processes (white noise vs.
noise with power low spectrum $f^{-k}$ with $k \geq 0$,
ordinal Brownian motion, fractional Brownian motion, and fractional Gaussian noise)
it can be  numerically ascertained that  {\it no\/}  forbidden patterns emerge.
Moreover, analytical expressions can be derived \cite{Shisha} for for some stochastic process
(i.e. fractional Brownian motion for PDF's based on ordinal patterns with length $2 \leq D \leq 4$).
%%%%%%%%%%%%%%%%%%%%%%%%%%%%%%%%%%

In the case of time series generated  by {\it unconstrained
stochastic process\/} (uncorrelated process) every ordinal pattern
has the same probability of appearance
\cite{Amigo2006,Amigo2007,Amigo2008,Amigo2010}. That is, if the
data set is long enough, all the ordinal patterns will eventually
appear. In this case, when the number of time-series' observations
becomes long enough the associated probability distribution
function should be the uniform distribution, and the number of
observed patterns will depend only on the length $N$ of the time
series under study.

For correlated stochastic  processes the probability of observing
individual pattern depends not only on the time series length $N$
but also on the correlation structure  \cite{Carpi2010}. The
existence of a non-observed ordinal pattern does not qualify it as
``forbidden'', only as {\it ``missing''\/}, and %is
%%%%%%%%%%%%%%%Martin-17-03-2011: is --->  it could be
it could be
%%%%%%%%%%%%%%%%%%%%%%%%%%%%%%%%%
due to the finite length of the time series.
A similar observation also holds
for the case of real data that always possess a stochastic
component due to the omnipresence of dynamical noise
\cite{Wold1938,Kurths1987,Cabanis1988}.
Thus, the existence of ``missing ordinal patterns" could be either
related to stochastic processes (correlated or uncorrelated) or to
deterministic noisy processes, which is the case for observational
time series.

Amig\'o and co-workers \cite{Amigo2006,Amigo2007}  proposed a test
that uses missing ordinal patterns to distinguish determinism
(chaos) from pure randomness in finite time series contaminated
with observational white noise (uncorrelated noise). The test is
based on two important practical properties: their finiteness and
noise contamination. These two properties are important because
finiteness produces missing patterns in a random sequence without
constrains, whereas noise blurs the difference between
deterministic and random time-series. The methodology proposed by
Amig\'o {\it et al.\/} \cite{Amigo2007} consists in a graphic
comparison between the decay of the missing ordinal patterns (of
length $D$) of the time series under analysis as a function of the
series length $N$, and the decay corresponding to white Gaussian
noise. This methodology was recently extended by Carpi {\it et
al.\/}  \cite{Carpi2010} for the analysis of missing ordinal
patterns in stochastic processes with different degrees of
correlation: fractional Brownian motion (fBm), fractional Gaussian
noise (fGn) and, noises with $f^{-k}$ power spectrum (PS) and ($k
\geq 0$). Specifically, this paper analyzes the decay rate of
missing ordinal patterns as a function of pattern-length $D$
(embedding dimension) and of time series length $N$. Results show
that for a fixed pattern length, the decay of missing ordinal
patters in stochastic processes depends not only on the series
length but also on their correlation structures. In other words,
missing ordinal patters are more persistent in the time series
with higher correlation structures. Ref. \cite{Carpi2010}  also
has shown that the standard deviation of the estimated decay rate
of missing ordinal patterns ($\alpha$) decreases with increasing
$D$. This is due to the fact that longer patterns contain more
temporal information and are therefore more effective in capturing
the dynamics of time series with correlation structures.

\section{Application to the logistic map with additive noise}
\label{sec:Logistica}

The logistic map constitutes  a canonic example, often employed to
illustrate new concepts and/or methods for the analysis of
dynamical systems. Thus, we will use the logistic map with
additive correlated noise in order to exemplify the behavior of
the normalized Shannon entropy ${\mathcal H}_S$ and the
Statistical Complexity ${\mathcal C}_{JS}$, both evaluated using a
PDF  based  on the Bandt-Pompe's procedure. We will also
investigate the decay rate of ``missing ordinal patterns" in both
a) the logistic map with additive noise and b) a pure-noise
series.

The logistic map is a polynomial  mapping of degree 2, $F: x_n
\rightarrow x_{n+1}$ \cite{Sprott2004}, described by the
ecologically motivated, dissipative system represented by the
first-order difference equation
\begin{equation}
x_{n+1} =  r \cdot x_n \cdot ( 1 - x_n )  \ ,
\label{logistica}
\end{equation}
with $0 \leq x_n \leq 1$ and $0 \leq r \leq 4$.

Let $\eta^{(k)}$ be a correlated noise with $f^{-k}$ power spectra
generated as describe in, for instance, \cite{Rosso2007}. The
steps to be followed are enumerated below.
\begin{enumerate}
%%%%%%Martin V20
\item Using the Mersenne twister generator \cite{mersenne1998} through the
$\textsc{Matlab}^\copyright$ {\sl RAND\/} function we generate
pseudo random numbers $y^0_i$ in the interval
$(-0.5,0.5)$ with an
{\it (a)\/} almost flat power spectra (PS),
{\it (b)\/} uniform PDF, and
{\it (c)\/} zero mean value.
\item Then, the Fast Fourier
Transform (FFT) ${y^1_i}$ is first obtained and then multiplied by
$f^{-k/2}$, yielding ${y^2_i}$;
\item  Now, ${y^2_i}$ is symmetrized so as to obtain a real function. The pertinent inverse FFT
is obtained, after discarding the small imaginary components produced by our numerical approximations.
\item  The resulting noisy time series is re-scaled to the interval $[-1,1]$, which
produces a new  time series $\eta^{(k)}$ that exhibits the desired
power spectra and, by construction,  is representative of
non-Gaussian noises.
\end{enumerate}

We considered a time series $\mathcal S = \{ S_n, n = 1, \cdots, N \}$ generated  by the discrete system:
\begin{equation}
S_n = x_n + A \cdot \eta^{(k)}_n \ ,
\label{series}
\end{equation}
in which, $x_n$ is given by the logistic map and  $\eta^{(k)}_n \in [-1,1]$ represents a noise
with power spectrum  $f^{-k}$ and amplitude $A$.

\begin{figure}
\noindent
\includegraphics[width=5in]{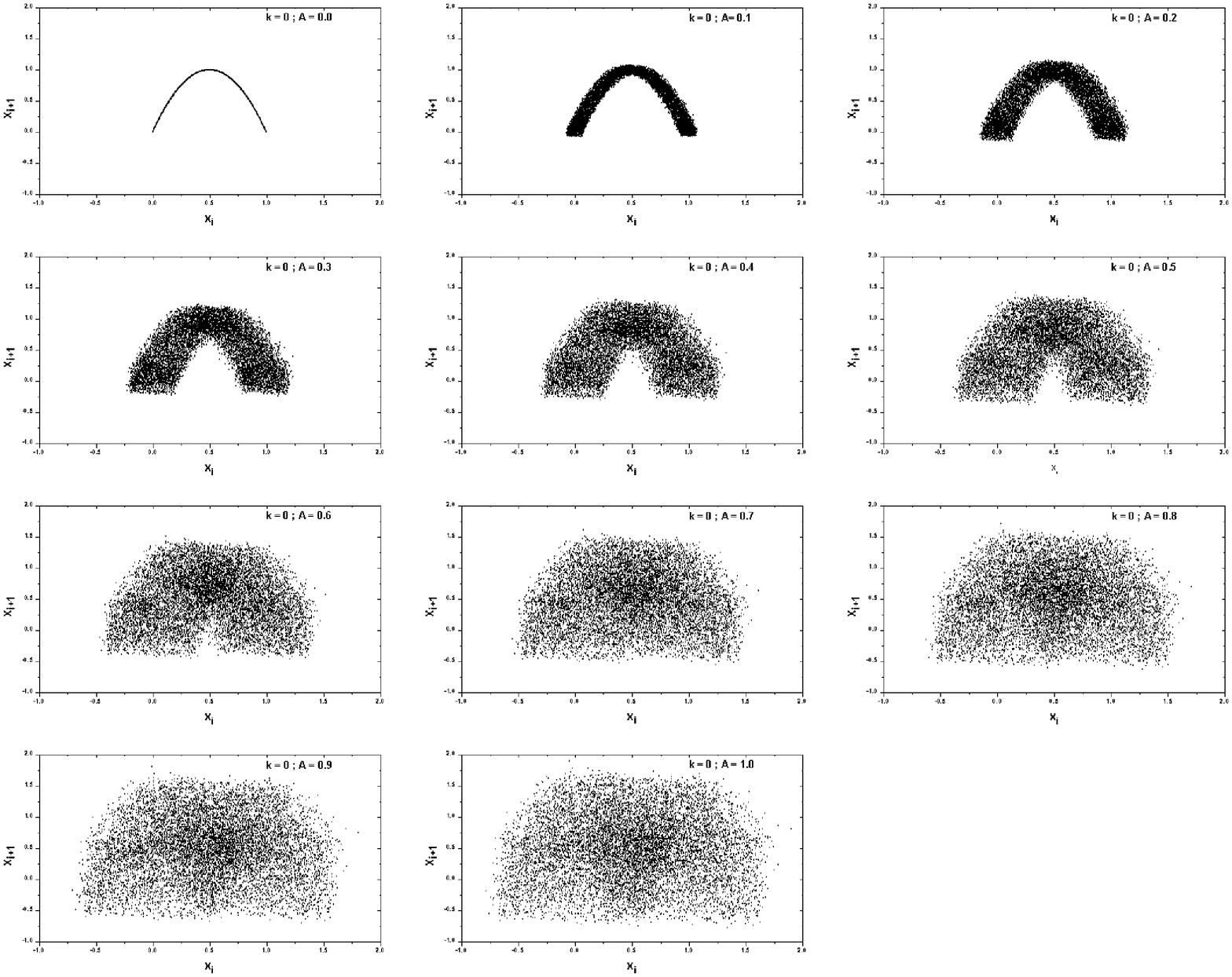}
\caption{ First return map corresponding to a typical signal from
a noise-contaminated logistic map ($r=4$). The additive noise has
a power spectrum of the $f^{-k}$-type ($N=10^4~data$).
We consider the case $k=0$ (white noise) with noise amplitude $0 \leq A \leq 1$.
}
\label{seniales-k0}
\end{figure}

\begin{figure}
\noindent
\includegraphics[width=5in]{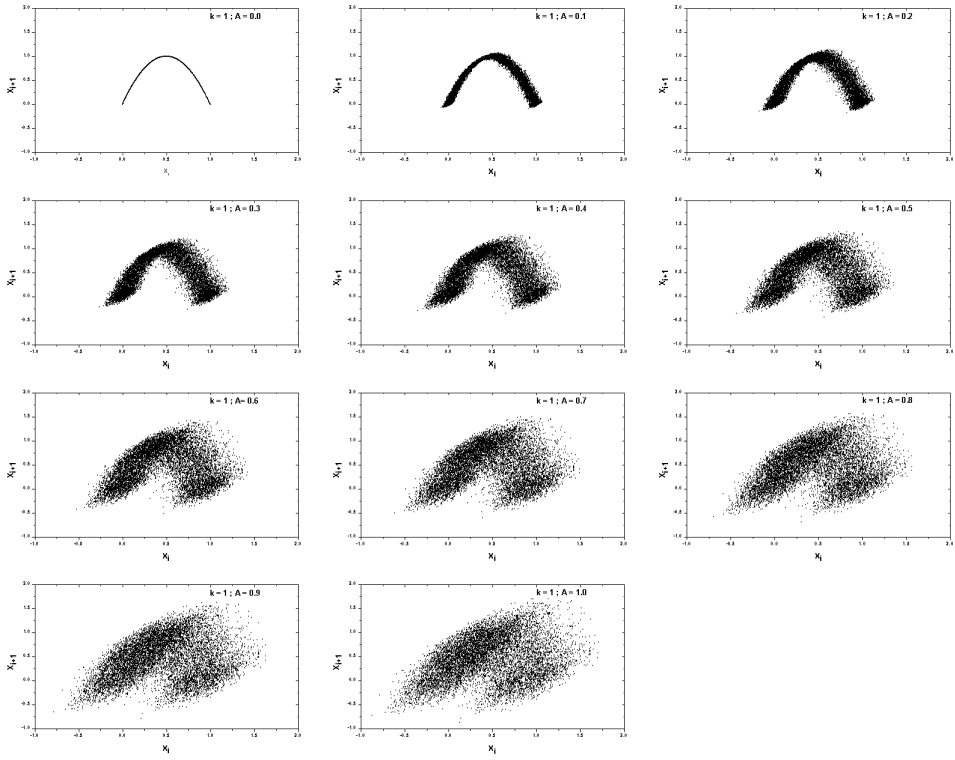}
\caption{ Same as in Fig. 1 but for $k=1$. }
\label{seniales-k1}
\end{figure}

In generating the logistic map's component of our time-series we
fixed $r=4$ and  started the iteration procedure with a random
initial condition. The first $5\cdot10^4$ iterations were
considered part of the transient behavior and were therefore
discarded. After the transient part died out, $N=10^5~data$ were
generated. In the case of the stochastic component of our
time-series we considered $0 \leq k \leq 2$ with $k$-values
changing by the amount $\Delta k = 0.25$. The noise amplitude was
varied between $0 \leq A \leq 1$ with increments $\Delta A = 0.1$.
Ten noise time series of length $N=10^5~data$ (different seeds)
were generated for each value of the $k$-exponent. Figures
\ref{seniales-k0} and \ref{seniales-k1} display the first return
map corresponding to a typical signal obtained from Eq.
(\ref{series}) (we show only the first $N=10^4~data$ points), for
the cases of $k=0$ (white noise) and $k=1$ respectively. These
graphs clearly show how the distorting effect of the additive
noise on the logistic time series increases as the amplitude $A$
grows.

\section{Results and discussion}
\label{sec:Resultados}

An important quantity for us,  called ${\mathcal M}(N,D)$, is the
average number of missing ordinal patterns of length $D$,
%%%%%%%%%%%%%%%%OSVALDO-17-03-2011: cambie la redaccion a pedido del Refere 1
that is, patterns which are
%%%%%%%%%%%%%%%%%%%%%%%%%%%%%%%%%%
{\it not\/} observed in a time series (TS) with $N$ data values. As we
mentioned before, for pure correlated stochastic processes the
probability of observing an individual pattern of length $D$
depends on the time series-length $N$ and on the correlation
structure (as determined by  the type of noise $k>0$ ). In fact,
as  shown for noises with an $f^{-k}$ Power Spectrum  in
Ref.~\cite{Carpi2010}, as the value of $k>0$ augments -- which
implies that correlations grow -- increasing values of $N$ are
needed if one wishes to satisfy the ``ideal" condition ${\mathcal
M}(N,D)=0$. If the time series is chaotic but has an additive
stochastic component, then one expects that as the time series'
length $N$ increases, the number of ``missing ordinal patterns"
will decrease and eventually vanish. That this may happen does not
only depend on  the length $N$  but  on the de
correlation-structure of the added noise as well.

\begin{figure}
\noindent
\includegraphics[width=5in]{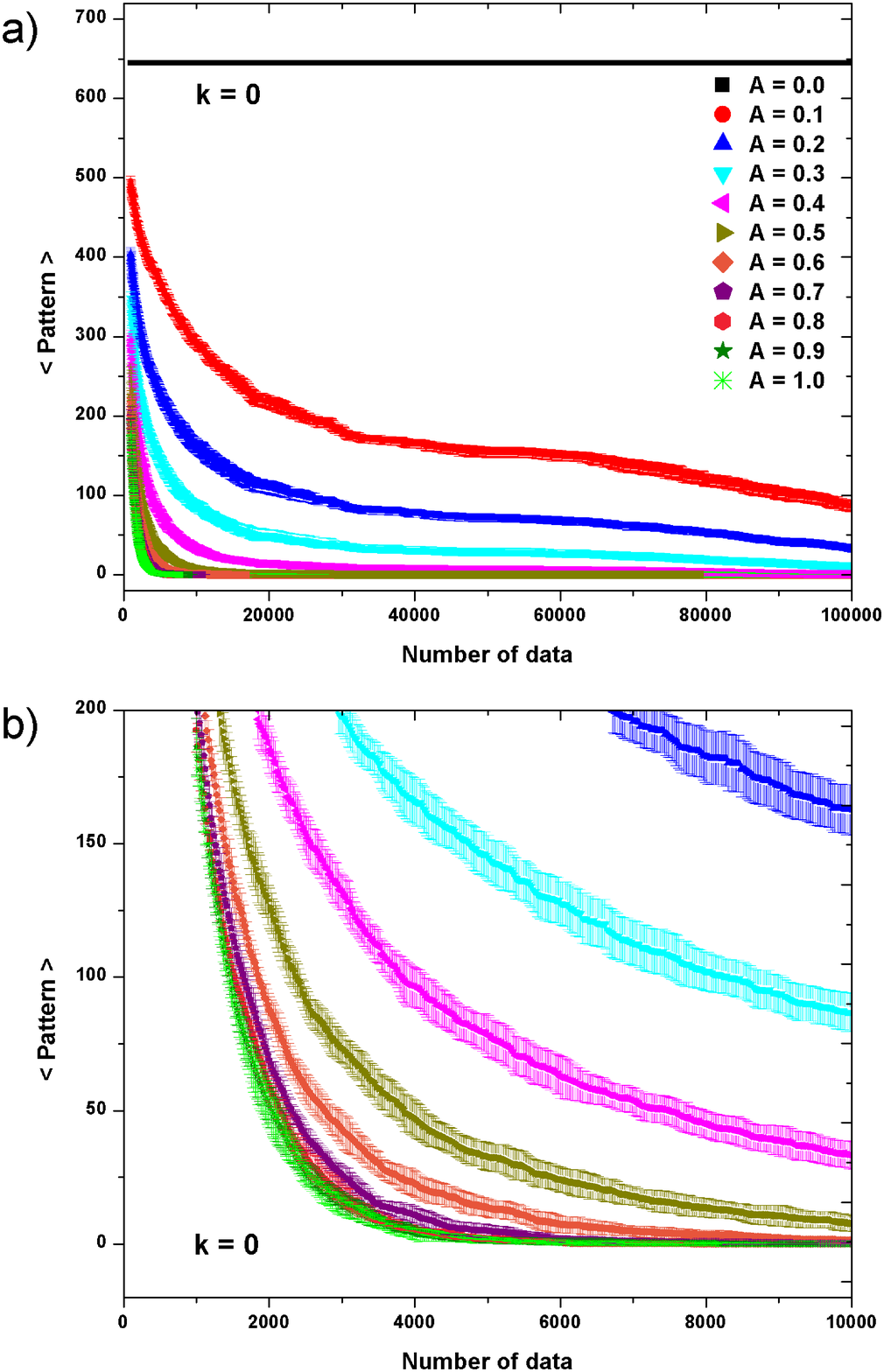}
\caption{ {\it a)\/} Average over  ten time series  of the missing
patterns number $<{\mathcal M}(N,D)>$ ($mean \pm SD$), as a
function of the time series length $10^3 \leq N \leq 10^5$, for
the case $k=0$ (withe noise) and $D=6$. Note that for the case
$A=0$ we have a purely logistic time series and the associated
number of missing patterns $<Pattern> = 645$ is independent of the
time series length. {\it b)\/} Details over $10^3 \leq N \leq 10^4$.}
\label{pattern-N-k0}
\end{figure}

Here from we will fix the pattern-length at $D=6$ and the time
series (TS) length at $N=10^5~data$. In Fig. \ref{pattern-N-k0}
the average number of missing patterns \newline $<{\mathcal
M}(N,D)>_{k=0}$ (mean value and standard deviation over  ten time
series (see Eq.~(\ref{series})) is displayed as a function of the
number of TS-data. ${\mathcal M}$ was evaluated as a function of
the TS-length $10^3 \leq N \leq 10^5$ with $\Delta N= 20$. $A=0$
corresponds to the case of the pure logistic time series and
consequently the number of missing patterns is constant
$<{\mathcal M}> = 645$ (see Fig. \ref{pattern-N-k0}.a). From this
plot we gather that the mean number of missing patterns rapidly
decreases as the noise amplitude increases, and could vanish when
the time series' length is large enough.

\begin{figure}
\noindent
\includegraphics[width=5in]{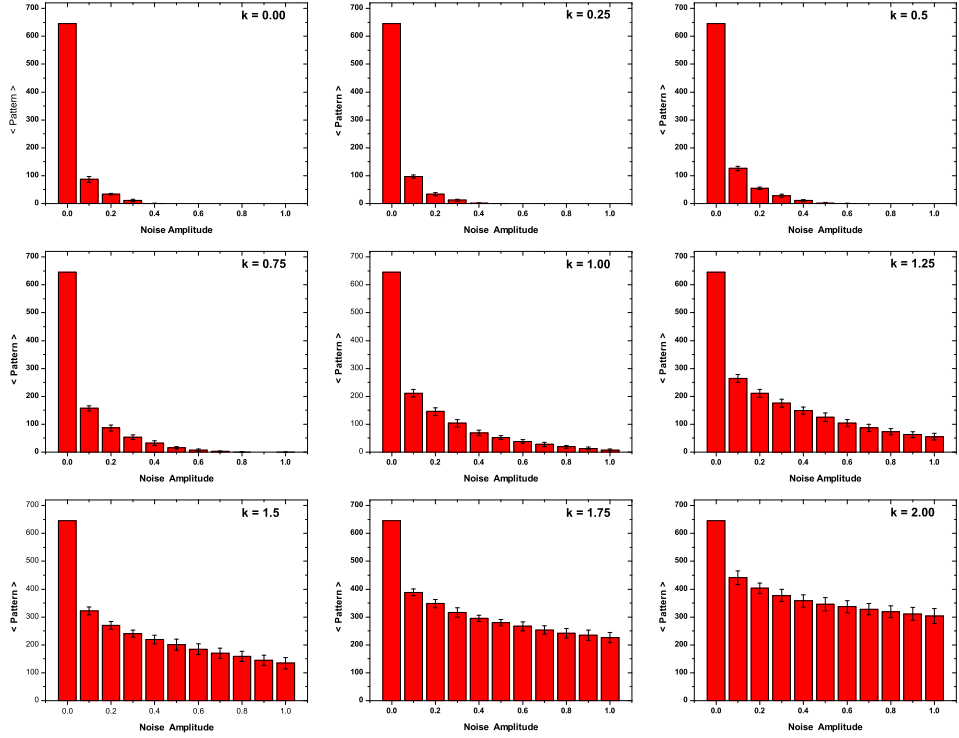}
\caption{ Average values corresponding  to the number of missing
patterns $<{\mathcal M}(N,D)>$ ($mean \pm SD$) for times series of
$N=10^5~data$ and $D=6$, as a function of the noise amplitude $A$
and for several  $k$ values. } \label{patrones-k}
\end{figure}

We pass now to  consider  the quantity $<{\mathcal M}(N,D)>_{k \ne
0}$ (see Eq.~[\ref{series})] corresponding to the TS-number of
missing patterns (mean value and standard deviation)   as a
function of the noise amplitude $A$. This is the purpose of  Fig.
\ref{patrones-k}. Due to the finiteness of the record length $N$
we find a non zero number of missing patterns. The noise
correlation effect is more significant for values of $k>1$. This
graph illustrates on the dependence of the number of missing
patterns on {\it i)\/} the noise amplitude $A$ and {\it ii)\/} the
noise-correlation $k$.

\begin{figure}
\noindent
\includegraphics[width=5in]{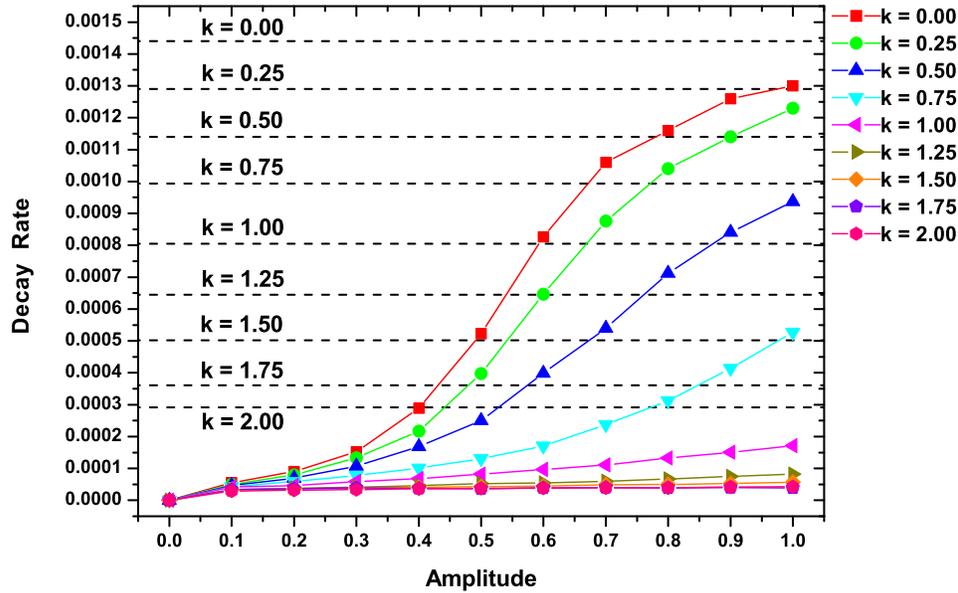}
\caption{ Decay rate $\alpha$ for the missing patterns of the
noise-contaminated logistic time series,
  for different values of $k$
as a function of the additive noise component $A$. Horizontal
lines correspond to the  pure noise-value with $f^{-k}$ power spectrum.
The times series' length is
%%%%%%%%%%%%%OSVADLDO-17-03-2011: Referee 3, minor 2
$10^3 \leq N \leq 10^5~data$
and the  pattern length used is $D=6$. }
\label{decay-amplitude}
\end{figure}

Let us now focus our attention on how the average missing ordinal
patterns $<{\mathcal M}(N,D)>$ decays as  $N$ grows (for a fixed
value of the pattern length $D$) by recourse to the  exponential
law advanced by Amig\'o {\it et al.} \cite{Amigo2007,Amigo2008}
and Carpi {\it et al.} \cite{Carpi2010}, namely,
\begin{equation}
\label{decay}
{\mathcal M}(N, D) =  B \cdot  \exp\{ - \alpha \cdot N \}  \ ,
\end{equation}
where $\alpha \geq 0$ is the characteristic decay rate and $B$ is
a constant.  $\alpha$ is determined by fitting Eq.~(\ref{decay})
to the numerically generated values of $<{\mathcal M}(N,D)>$ (for
each $k$ and $A$) using the least square method. Figure
\ref{decay-amplitude} displays the $\alpha-$values  for different
$k$-values as a function of the noise amplitude $A$. In the same
plot we also represent, as horizontal lines, the corresponding
values of the decay rate $\alpha$ pertaining to an strictly noisy
TS. In all cases, for the evaluation of ${\mathcal M}(N,D)$ we
consider $10^3 \leq N \leq 10^5$, with $\Delta N= 20$ and $D=6$.
The mean value of the $r^2$ coefficient of the exponential fit
(the ``goodness" of the fit)  was always larger than $0.95$.

We see in Fig. \ref{decay-amplitude} that, if $A=0$ (logistic
series with no noise), the decay rate vanishes.  For $A \neq 0$
 strictly forbidden patterns do not exist ($\alpha
>0$). {\it Missing patterns} are those that had not appeared before due to the finite
TS-length' effect. When the amplitude of the added noise
increases, the number of these missing patterns tends to decrease.
However, their observation-probability depends on the time series'
length ($N$) and on the noise-correlation ($k$).

We can regard the additive noise as a perturbation to the logistic
data (see Figs. \ref{seniales-k0} and \ref{seniales-k1}). As a
result of this perturbation the original patterns can change and
so does the PDF associated to the TS. The number of observed
patterns becomes now  a combination of those originated by,
respectively, the deterministic and the stochastic series'
components (their interaction may play a role as well).
Note that the added noise can destroy true forbidden patterns, although it
can create also new forbidden patterns.
It is clear from Figs. \ref{patrones-k} and \ref{decay-amplitude} that, for a fixed
number of data, the number of missing patterns  tends to decrease
and the corresponding decay rate $\alpha >0$,
%%%%%%%%%%%%%%%%OSVALDO-17-03-2011: Referee 3, punto 4, tiene razon.
%%%%%%%%%%%%%%%%%%%%%%%%%%%%%%%%%%% Cambio: tend to grow ---> present smaller values
% tends to grow
displays smaller values
%%%%%%%%%%%%%%%%%%%%%%%%%%%%%%%%%%
if both the noise amplitude $A$ and the correlation $k$ increase.
However, as expected, for any given value of $k$, the logistic series
contaminated with noise has a lower decay rate than that of the
strictly noisy one.
If the logistic map is contaminated with non correlated noise
($k=0$) then, as Fig.~\ref{decay-amplitude} tells us, $\alpha <
\alpha^{(k=0)}$ for all noise amplitudes ($0 < A \le 1$) and
deterministic influences on the TS can be established via the
distance from the  $\alpha$-decay curve to the strictly stochastic
horizontal lines.
%%%%%%%%%%%%%%%%OSVALDO-17-03-2011: Referee 3, punto 5, tiene razon, no es claro que sea asi, ergo borre la frace.
%These distances diminish as the  the noise becomes more correlated.

%%%%%%%%%%%%%%%%OSVALDO-17-03-2011: respuesta al segundo comentario del Referee 1.
%%%%%%%%%%%%%%%%%%%%%%%%%%%%%%%%%%% Tiene razon, para que no haya ambiguedad, uno deberia tener una
%%%%%%%%%%%%%%%%%%%%%%%%%%%%%%%%%%% estimacion del k para el ruido.
%In a real situation, of course, the
%strength of the contaminating noise and  its degree of correlation
%are both unknown.
%On the basis of our results we conjecture that:
In a real situation, of course, the strength of the contaminating
noise is un-known. However, were we are able to forecast an
estimation or at least make an educated guess of the power factor
$k$ of the contaminating noise, 
%%%%%%%%%%%%%%%%%%%%%%%%%%%%%%%%%%
our results would motivate us to conjecture that:
\begin{description}
  \item[$\bullet$] Given an observational
  time series of length $N$ and missing patterns decay rate $\alpha^{(TS)}$, and
  \item[$\bullet$] a pure $k$-noise time series of the same length $N$, with $\alpha^{(k-noise)}$,
  \item[$\bullet$] if $\alpha^{(TS)} < \alpha^{(k-noise)}$,
  then the observational time series can be associated to some deterministic
  process with an additive noise component.
\end{description}

Summing up,  if one employs for  the TS-analysis {\it i)\/}
quantifiers based on Information Theory and {\it ii)\/} the Bandt
and Pompe's PDF, several  effects have to be considered, namely,
those associated to the
    {\it (a)\/} finite number of data $N$;
    {\it (b)\/} noise contamination level (represented by $A$);
    {\it (c)\/} noise correlation (represented by $k$); and
    {\it (d)\/} persistence of forbidden patterns linked to the deterministic component of the series.

The missing patterns' quantifier $\alpha$ can be used for
practical proposes in order to characterize an observational time
series. However, given the nature of the  above mentioned effects,
specific quantifiers like ``entropy"  ${\mathcal H}_S$ and
``complexity"
  ${\mathcal C}_{JS}$ could potentially reveal {\it additional} important aspects associated
to the series-PDF that can not be discerned via the behavior of
$\alpha$. We pass now to a discussion centered on the two just
mentioned  quantifiers.

\begin{figure}
\noindent
\includegraphics[width=5in]{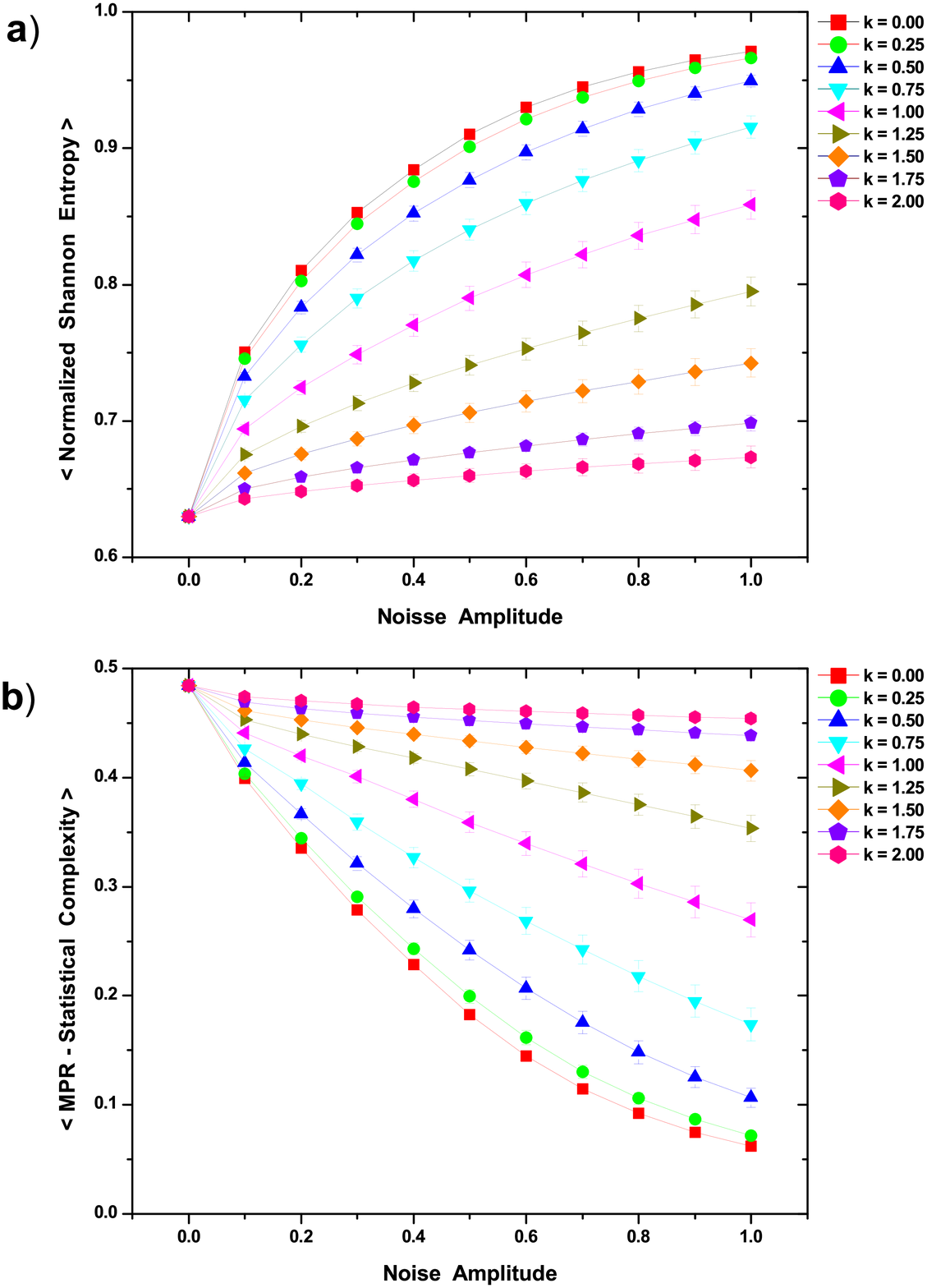}
\caption{ Average values ($mean \pm SD$) corresponding to {\it
(a)\/} the normalized Shannon entropy ${\mathcal H}_S$ and {\it
(b)\/} the statistical complexity ${\mathcal C}_{JS}$, for times
series of $N=10^5~data$ and $D=6$, as a function of the noise
amplitude $A$ and for several  $k$ values. } \label{H-C-k}
\end{figure}

For each one of the ten time-series generated for each pair
($k$,$A$),  the normalized Shannon entropy ${\mathcal H}_S$ and
MPR statistical complexity measure ${\mathcal C}_{JS}$ were
evaluated using i) the Bandt and Pompe PDF and ii)  the procedure
described in previous sections. Fig. \ref{H-C-k} displays the
pertinent values for series of length $N=10^5~data$ and $D=6$. If
$k \approx 0$ we see from these graphs that, for increasing values
of the amplitude $A$, entropy and complexity values change
(starting from the value corresponding to the pure logistic
series, i.e., $A=0$) with a tendency to approach the values
corresponding to pure noise, that is, (${\mathcal H}_S \approx 1$
and ${\mathcal C}_{JS} \approx 0$). As $k$ increases, the
dependence on the noise-amplitude $A$ tends to become attenuated,
and it almost disappears for $k \approx 2$. This is due to the
effect of the ``coloring" (correlations) of the noise, that grows
with increasing values of $k$.

\begin{figure}
\noindent
\includegraphics[width=5in]{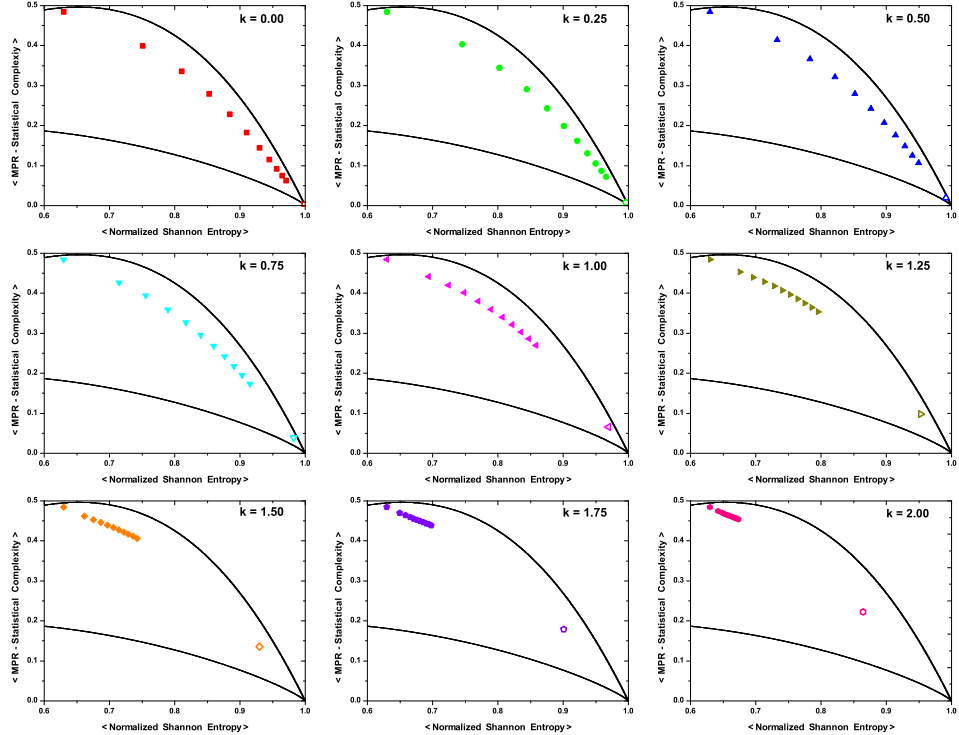}
\caption{ The causality entropy-complexity plane associated to
Fig. \ref{H-C-k} for  different values of both $k$ and  $A$ (times
series length $N=10^5~data$ and $D=6$). The corresponding values
for a purely noisy series ($N=10^5~data$ and $D=6$) are also displayed in
the guise of open symbols. The continuous lines represent the
values of minimum and maximum statistical complexity ${\mathcal
C}_{min}$ and ${\mathcal C}_{max}$, evaluated for the case of
pattern length $D=6$. In each graph,  the first point at the west
has $A=0$ (to the left) and the last at the east $A=1$ (to the
right). }
\label{HxC-k}
\end{figure}

Using the  values displayed  in  Fig. \ref{H-C-k} we generate the
associated causality entropy-complexity plane for each one of the
relevant $k$ values  and for  different noise-amplitudes $A$.
These planes are depicted in Fig. \ref{HxC-k}. We also display in
these planes  values pertaining to  the  pure-noise time series
(with the same record length $N=10^5~data$) in the guise of  open
symbols. The continuous lines represent the values of minimum and
maximum statistical complexity ${\mathcal C}_{min}$ and ${\mathcal
C}_{max}$, respectively,  evaluated for $D=6$.

Of course, for $A=0$ we obtain the purely deterministic value
corresponding to the logistic time-series, localized approximately
at a medium-hight entropic coordinate,   near the highest possible
complexity. Note that such is the typical behavior observed for
deterministic systems \cite{Rosso2007}. For a purely uncorrelated
stochastic process ($k=0$) we have ${\mathcal H}_S = 1$ and
${\mathcal C}_{JS} = 0$. The correlated (colored) stochastic
processes ($k \neq 0$) yield points located at intermediate values
between the curves ${\mathcal C}_{min}$ and ${\mathcal C}_{max}$,
with decreasing values of entropy and increasing values of
complexity as $k$ grows \cite{Rosso2007} (see Fig. \ref{H-C-k}).

Fig. \ref{HxC-k} shows  that the main effect of the additive noise
($A \neq 0$) is to shift  the point representative of  zero noise
($A=0$) towards increasing values of entropy ${\mathcal H}_S$ and
decreasing values of complexity ${\mathcal C}_{JS}$. This shift
defines a curve that is located in the vicinity of the maximum
complexity ${\mathcal C}_{max}-$curve. Such behavior can be linked
to the persistence of  forbidden patterns because they, in turn,
imply that the deterministic logistic component is still operative
and  influencing the TS-behavior.

We stress the fact that noise acts here as an additive
perturbation to the logistic series. We may assert that some
degree  of robustness of our (causality) entropy-complexity plane against %(ECP)
the effects of TS-noise contamination becomes evident. Moreover,
a clear differentiation can be appreciated between  the
localization of the points corresponding to the time series with
additive noise with  those from a ``pure-noise" series.
We insist on the fact that such entropy-complexity plane's results %ECP-results
are in agreement  with the  persistence of forbidden patterns from
the deterministic component of the time-series.

\begin{figure}
\noindent
\includegraphics[width=5in]{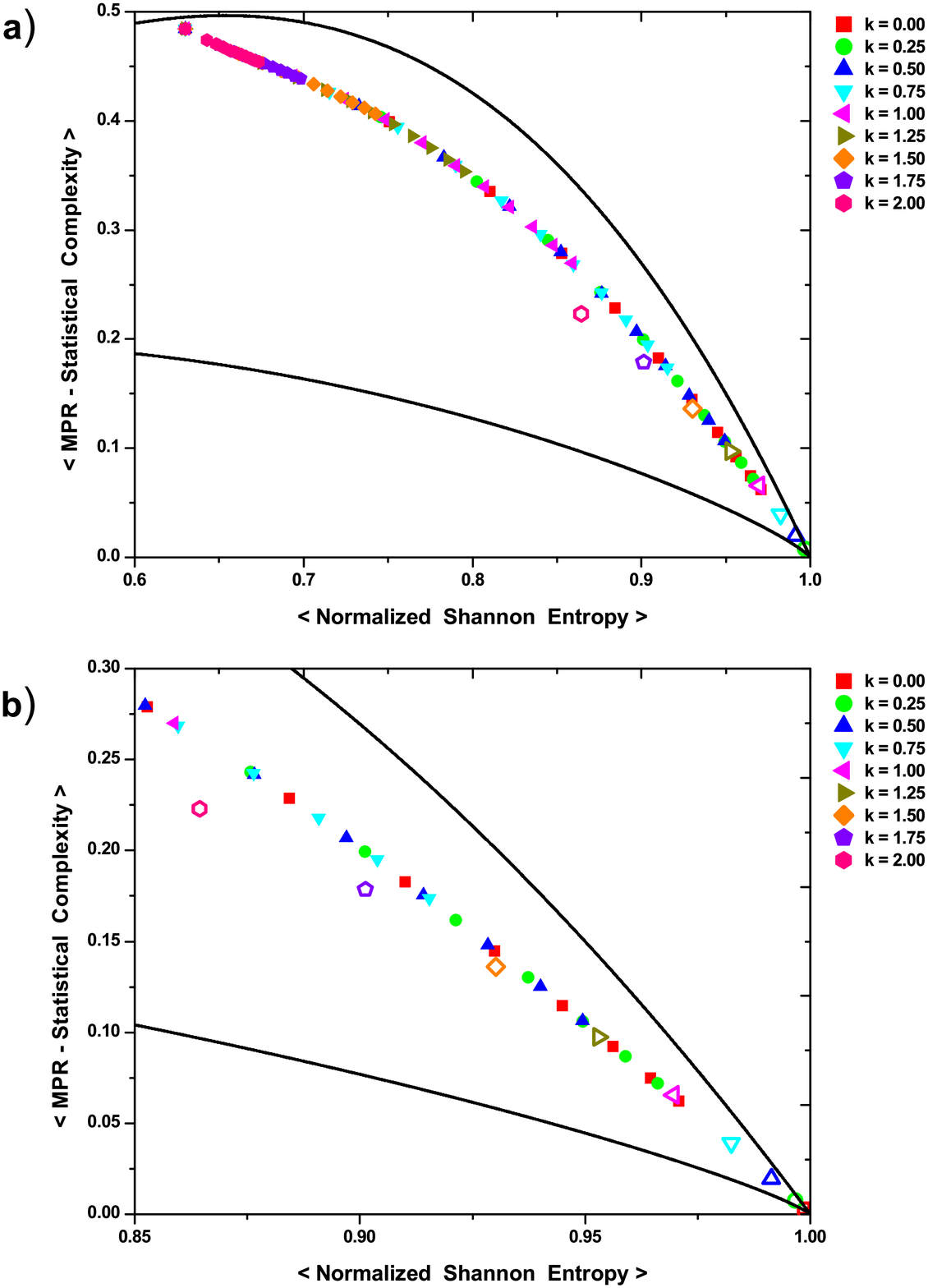}
\caption{ {\it (a)\/} The causality entropy-complexity plane for
all values displayed in (Fig. \ref{H-C-k}) corresponding to
different of both $k$  and  $A$ (times series length $N=10^5~data$
and $D=6$). Values for the case of strictly noisy time series
($N=10^5~data$ and $D=6$) are also shown as open symbols. The
continuous lines represent the values of minimum and maximum
statistical complexity ${\mathcal C}_{min}$ and ${\mathcal
C}_{max}$ evaluated for the case of pattern length $D=6$. {\it
(b)\/} Amplification. } \label{HxC-k-todos}
\end{figure}

Interestingly enough, if we plot  these values (for all $A$) in a
single graph, as in Fig. \ref{HxC-k-todos}, we clearly appreciate
the behavior described in the preceding paragraphs. Moreover, all
the pertinent  points yield  a curve that closely approaches the
maximum complexity one ${\mathcal C}_{max}$. We can hypothesize
that this curve is characteristic of the dynamical deterministic
system, and could be used in order to determine the  level of
noise-contamination. Research  in this direction is being
currently undertaken.

\section{Conclusions}
\label{sec:Conclusions}

We have looked carefully at the concept of forbidden/missing
ordinal patterns. Why? Because it has been recently used as a tool
for the discrimination between deterministic and stochastic
behavior in observational time series
\cite{Amigo2006,Amigo2007,Amigo2008,Zunino2009,Ouyang2009,Zanin2008}.
Given the importance of the subject, that in the light of these
previous efforts becomes evident, in the present work we have
extended the analysis of  missing ordinal patterns
\cite{Amigo2010,Carpi2010} and linked it to the causality
entropy-complexity plane \cite{Rosso2007}. Our considerations were
made  with regards to deterministic finite time series
contaminated (as a perturbation) with additive noises of different
degree of correlation.

We were able to show that, by comparing the decay rate of missing
ordinal patterns both in a noisy time series and in a pure-noise
(with some degree of correlation) one, insights pertaining to the
deterministic component of the original time series can be gained.

This seemingly worthy goal  was  achieved, without the need of
additional numerical work such as the one needed for evaluating the decay rate of missing
ordinal patterns.
It just suffices to  look carefully  at the planar locations of the pertinent points,
for each time series. Moreover, we have encountered a
characteristic behavior that can be associated to variations in
the amount of noise-contamination which, in turn, facilitates the
identification of the time-series' deterministic component.

\section*{Acknowledgments}
This research has been partially supported by a scholarship from The University of Newcastle awarded to Laura C. Carpi.
Osvaldo A. Rosso gratefully acknowledges support from CAPES, PVE fellowship, Brazil.
Mart\'{\i}n G\'omez Ravetti acknowledges support from FAPEMIG and CNPq, Brazil.

\newpage
%FIGURA 1: Time Series - k=0

\newpage
%FIGURA 2: Time Series - k=1

\newpage
%FIGURA 3: Missing patterns vs N - K=0

\newpage
%FIGURA 4: Missing patterns vs A for k and N=10^5

\newpage
%FIGURA 5: Decay Rate vs Amplitude para N=10^5

\newpage
%FIGURA 6: H and C vs A for k and N=10^5

\newpage
%FIGURA 7:Plano HxC for k , A and N=10^5

\newpage
%FIGURA 8:Plano HxC for k , A and N=10^5

\end{document}